\documentstyle[12pt,dina4,preprint,psfig,aps]{revtex}
\begin{document}
\draft
\tightenlines

\title{ Are we close to the QGP? --
	Hadrochemical vs. microscopic analysis of particle production in 
	ultrarelativistic heavy ion collisions. 
\footnote{supported by GSI, BMBF, Graduiertenkolleg ``Theoretische
	und experimentelle Schwerionenphysik''  and DFG} }

\author{
        S.~A.~Bass, M.~Belkacem, M.~Brandstetter,
	M.~Bleicher, L.~Gerland, J.~Konopka, L.~Neise, C.~Spieles,
	S.~Soff, H.~Weber,
	H.~St\"ocker and W.~Greiner}
\address{
  	Institut f\"ur Theoretische Physik \\
	Johann Wolfgang Goethe Universit\"at\\
	Robert Mayer Str. 8-10\\
	D-60054 Frankfurt am Main, Germany
}

\maketitle

\begin{abstract}
Ratios of hadronic abundances 
are analyzed for pp and nucleus-nucleus 
collisions at $\sqrt{s} \approx 20$ GeV 
using the microscopic transport model UrQMD.
Secondary interactions significantly change  the primordial 
hadronic cocktail of the system. A comparison
to data shows a strong dependence
on rapidity. Without assuming
thermal and chemical equilibrium,
predicted hadron yields and ratios 
agree with many of the data
($\pi/p$, $d/p$, $\bar p/p$, $\bar \Lambda/\Lambda$, $\bar \Xi/ \bar \Lambda$ etc.). 
Large discrepancies
to the data ($> 50$ \%) are found for the
$K^0_S/\Lambda$ and $\Omega/\Xi$  ratios. 
\end{abstract}

\pagebreak

Hadron abundances and ratios
have been suggested as possible signatures  
for exotic states and phase transitions in dense
nuclear matter.
In addition they have been applied
to study the degree of chemical equilibration in a relativistic
heavy-ion reaction.
Bulk properties like temperatures, entropies and chemical potentials
of highly excited hadronic matter have been extracted
assuming thermal and chemical equilibrium
\cite{stoecker,stock86a,rafelski,rafelski2,cleymans,braun-munzinger,spieles97a}.

The present work confronts the conclusions of a series of publications
which have attempted to fit the available 
AGS 
\cite{exis_ags}
and SPS 
\cite{exis_sps} 
data on hadron yields
and ratios. The latter have been done
either in the framework of a hadronizing QGP droplet 
\cite{spieles97a,barz88a}
or of a hadron gas in
thermal and chemical equilibrium \cite{braun-munzinger}.
(including elementary proton-proton interactions \cite{becattini97a}).
It has been shown that the thermodynamic parameters $T$ and $\mu_B$
imply that these systems have been either very close to 
or even above the critical
$T$, $\mu_B$ line for QGP formation \cite{braun-munzinger,spieles97a}.

Here, in contrast, the non-equilibrium
microscopic Ultra-relativistic Quantum Molecular Dynamics 
transport model (UrQMD) \cite{uqmdref} 
is used to calculate hadron ratios 
without thermalization assumptions.
We tackle the following questions:
\begin{enumerate}
\item Is this microscopic model able to predict elementary hadron production
(including yields and ratios)?
\item  How do hadron ratios in  elementary nucleon-nucleon
interactions compare to those stemming from the final state of a 
nucleus-nucleus reaction? 
\item Do isospin and secondary interactions (rescattering)
play a major role or is the hadronic makeup of the system fixed after
the first primordial highly energetic nucleon-nucleon collisions?
\item To what extent do the hadron ratios depend on rapidity and
transverse momentum? How strong is their sensitivity to experimental
acceptance cuts?
\end{enumerate}

The UrQMD model \cite{uqmdref}
is based on analogous principles as 
(Relativistic) Quantum Molecular Dynamics
\cite{peilert88a,hartnack89b,aichelin91a,sorge89a,lehmann95a}.
Hadrons are represented by Gaussians   
in phase space.
The nucleons  
are initialized in spheres of radius
$R = 1.12 A^{1/3}$ fm. Momenta are chosen 
according to a non-interacting Fermi-gas ansatz. 
Each nucleon occupies
a volume of $h^3$, thus phase space is uniformly
filled (in a statistical sense). 
Hadrons are then propagated according
to Hamilton's equation of motion.
The microscopic evolution of the hadrochemistry in
heavy-ion reactions
requires the solution of a set of hundreds of coupled (Boltzmann-type)
integro-differential equations.
This means that 
all (known) hadrons need to be included into the model as realistically
as possible.
The collision term of the UrQMD model treats 55 different
isospin (T) degenerate baryon (B) species
(including nucleon-, delta- and hyperon- resonances with masses up to 2 GeV) 
and 32 different T-degenerate meson (M) species, 
including (strange) meson resonances as well as the 
corresponding anti-particles, i.e. 
full baryon/antibaryon symmetry is included.
Isospin is explicitely treated (although the SU(2) multipletts are
assumed to be degenerate in mass).
For excitations with masses $>2$ GeV (B) and 1.5 GeV (M) 
a string model is used.
All (anti-)particle states can be produced -- in accord with the conservation
laws -- both, in the string decays as well as in s-channel
collisions or in resonance decays.

Tabulated or parameterized experimental cross sections are used when 
available. Resonance absorption and scattering is handled via the 
principle of detailed balance. If no experimental information is
available, the cross section is either  calculated via
an OBE model or via a modified additive quark model,
which takes basic phase space properties into account.
The baryon-antibaryon annihilation cross section is parameterized as
the proton-antiproton annihilation cross section and then rescaled
to equivalent relative momenta in the incoming channel. 
Changes in the order of 50\% are observed if the same $\sqrt{s}$ dependence
is chosen for all baryon-antibaryon reactions.
For a detailed overview of the elementary cross sections and string excitation
scheme included in the UrQMD model, see ref. \cite{uqmdref}.

The UrQMD model allows for systematic studies of heavy-ion collisions over a
wide range of energies in a unique way: the basic concepts and the physics
input used in the calculation are the same for all energies. A relativistic
cascade is applicable over the entire range of energies from 100 MeV/nucleon
up to 200 GeV/nucleon (a molecular dynamics scheme using a 
hard Skyrme interaction is used between 100 MeV/nucleon and 4 GeV/nucleon).
All calculations presented here have been performed
with UrQMD version 1.0 \cite{uqmdref}.

\begin{figure}
\centerline{\psfig{figure=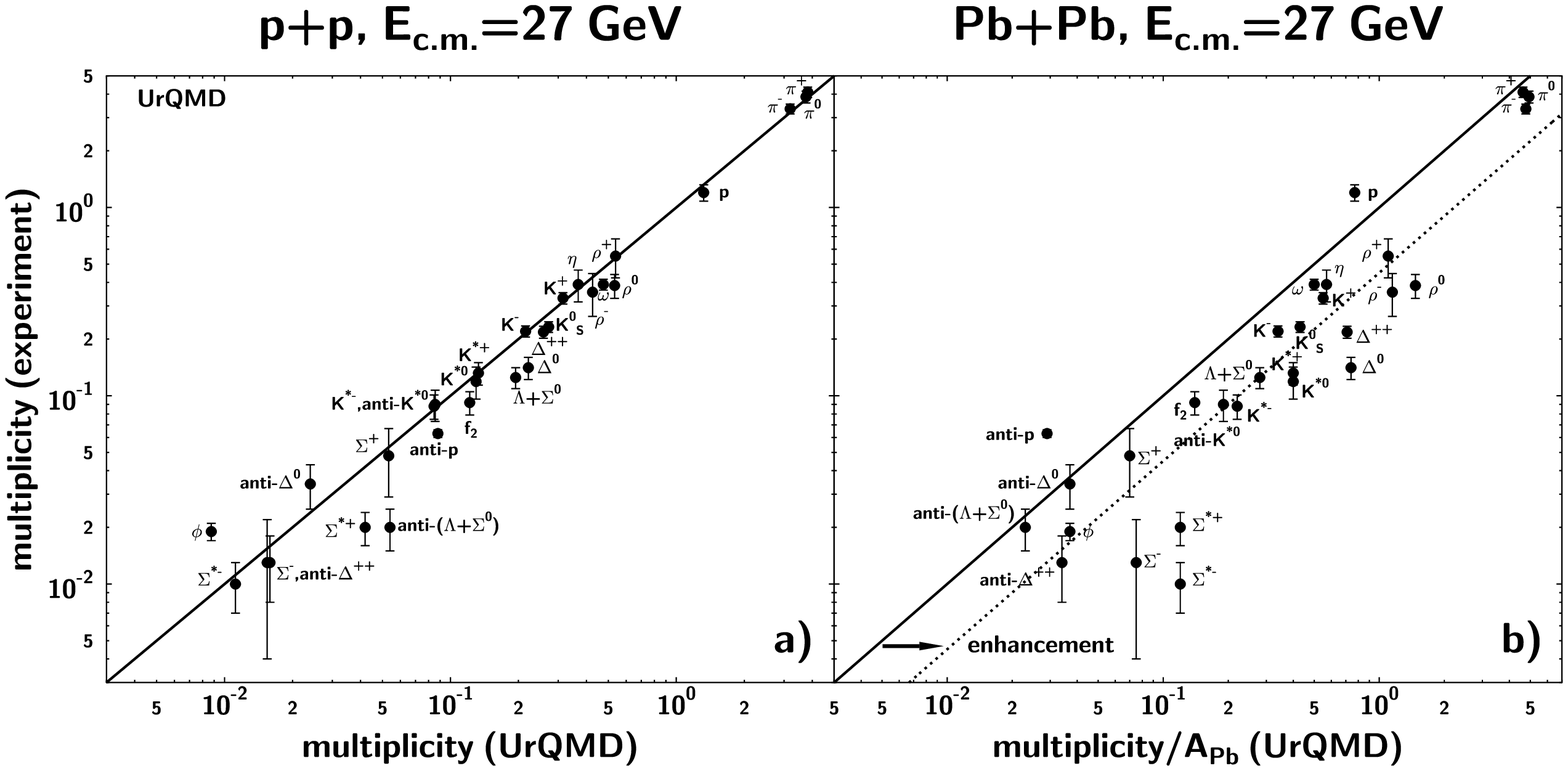,width=17cm}}
\caption{\label{becattini} Left: UrQMD hadron yields in elementary
proton-proton reactions at $ \protect\sqrt{s} =27 $ GeV compared to
data. The overall agreement spanning three orders of magnitude
is  good -- the most prominent deviations from the experiment
occur for the $\phi$-meson and for (anti-)$\Lambda + \Sigma^0$. 
Right: UrQMD hadron yields in central Pb+Pb reactions {\em per participating nucleon pair} 
at the same energy 
versus proton-proton data. The large shift to the r.h.s.  from the full diagonal line to
the dotted diagonal line marks the 
enhancement of secondary particle production in $AA$ reactions.
}
\end{figure}

Coming to the first question:
Is a microscopic model able to predict elementary hadron production
(including yields and ratios)?
Let us start with a comparison of 
a compilation of experimental measurements \cite{becattini97a} of
hadron production in elementary proton-proton collisions 
with yields as predicted by the UrQMD model in figure~\ref{becattini}a).
Note the overall good agreement (compatible to thermal model fits
\cite{becattini97a} yielding a temperature of 170 MeV) which spans
three orders of magnitude.
$\phi$-production
is underestimated by a factor of 2. $\Lambda + \Sigma^0$ 
(as well as the $\bar \Lambda + \bar \Sigma^0$) production is overestimated.
Problems in the strangeness sector are common to most string models
and indicate that strangeness production is not yet fully understood
on the elementary level \cite{pop95a}.
These deviations in the elementary channel 
have to be considered when comparing 
with heavy-ion experiments.

Unlike fireball models, UrQMD describes also
the momentum distributions (e.g. the d$N/$d$y$, d$N/$d$x_F$ and
d$N/$d$p_t$ distributions) for all hadron species under consideration. 
A detailed description and a
comparison to available hadron-hadron data can
be found in refs. \cite{uqmdref}.

\begin{figure}
\centerline{\psfig{figure=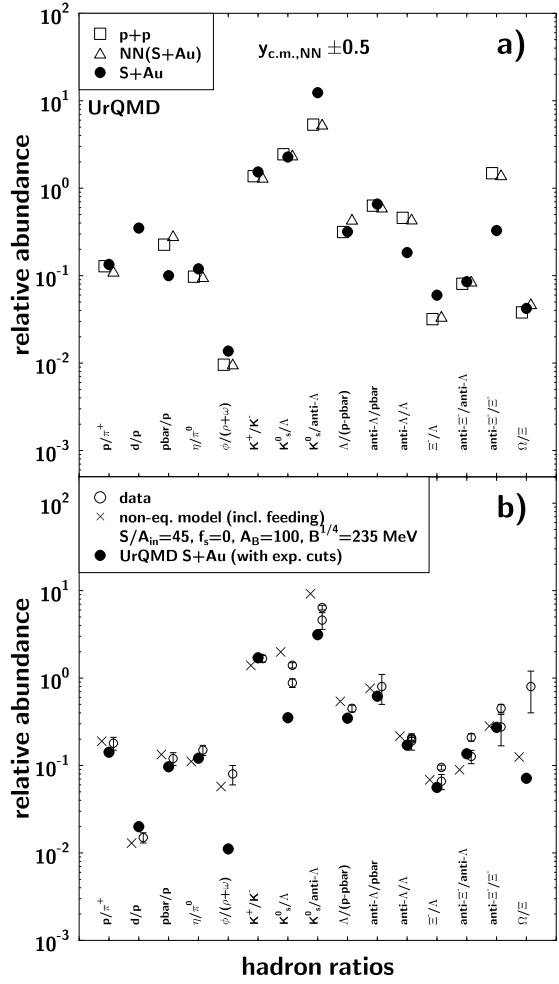,height=14cm}}
\caption{\label{sau_ratios} 
Top: UrQMD calculation of hadron ratios in S+Au
collisions at midrapidity (full circles). The ratios are compared to a
proton-proton calculation (open squares) and a nucleon-nucleon calculation
(correct isospin weighting) (open triangles).
Bottom: Comparison between the UrQMD model (full circles)
and data (open circles) 
for the system S+Au(W,Pb) at 200 GeV/nucleon. Also shown
is a fit by a microscopic hadronization model (crosses). 
Both non-equilibrium  models agree well with the data. Discrepancies are
visible for the $\phi/(\rho + \omega)$,
$K^0_S/\Lambda$ and $\Omega/\Xi$  ratios.
}
\end{figure}

Second: How do hadron ratios in  elementary nucleon-nucleon
interactions compare to those stemming from the final state of a 
nucleus-nucleus reaction? Do isospin and secondary interactions
play a major role or is the hadronic makeup of the system fixed after
the first primordial highly energetic nucleon-nucleon collisions?
Since even the particle abundances in elementary proton-proton
reactions may be described in a thermal model \cite{becattini97a}
one could speculate that the hadronic final state of a nucleus-nucleus
collision should not differ considerably from the primordial
``thermal'' composition.
The upper frame of figure~\ref{sau_ratios} 
shows hadron ratios calculated by the UrQMD
model for the S+Au system at CERN/SPS energies 
around midrapidity $y_{lab} = 3 \pm 0.5$ (full circles).
The ratios are compared to those stemming from a
proton-proton calculation (open squares) 
and from a nucleon-nucleon calculation, i.e. 
with the correct isospin weighting (open triangles)
for the primordial S+Au system, which is obtained by
weighting a cocktail of $pp$, $pn$ and $nn$ events in the following way:
$NN(S+Au)= 0.188 \cdot pp + 0.55 \cdot pn + 0.27 \cdot nn$.

The correct isospin treatment is of utmost importance, as it has 
a large influence on the primordial 
hadron ratios: 
Due to isospin conservation the $\bar p/p$ and
$\Lambda/(p - \bar p)$ ratios are enhanced by $\sim 30$\% and $\sim 40$\%, respectively;
it is easier
to produce neutral or negatively charged particles in a 
$nn$ or $pn$ collision than
in a $pp$ interaction. 

Rescattering effects,
which are visible when comparing the nucleon-nucleon
calculation (open triangles) with the full S+Au calculation (full circles),
have even a larger influence on the hadron ratios than isospin:
Changes in the ratios due to rescattering are easily 
on the order of 20\%-50\%.
Ratios involving antibaryons
even change by factors of $3-5$, due to their high hadronic annihilation
cross section. Most prominent examples are the ratios of
 $\bar \Xi / \Xi$ (factor 5 suppression),
 $\bar p/p$ (factor 3 suppression), 
$\bar \Lambda / \Lambda$ (factor 2 suppression),
$\Xi^-/\Lambda$ (factor 2 enhancement)
and $K^0_S/\bar \Lambda$ (factor 3 enhancement).

The lower frame of 
figure~\ref{sau_ratios} compares the UrQMD hadron ratios 
with experimental measurements 
\cite{exis_sps}.  
We use a data compilation which has been published in ref. 
\cite{braun-munzinger}.
The open circles represent the measurements whereas the full circles
show the respective UrQMD calculation for S+Au at 200 GeV/nucleon and 
impact parameters between 0 and 1.5 fm. For each ratio, the respective
acceptance cuts, as listed in \cite{braun-munzinger}, have been
applied. 
The size of the statistical errorbars of the UrQMD model 
does not exceed the size of the plot-symbols.
The crosses denote a fit with
a dynamical hadronization scheme, 
where thermodynamic equilibrium between a quark blob and the hadron 
layer is imposed \cite{spieles97a}. 
A good overall agreement between the data and the UrQMD
model is observed, of similar quality as that of the hadronization model.
Large differences between UrQMD and experiment, however, are visible in 
the $\phi/(\rho + \omega)$,
$K^0_S/\Lambda$ and $\Omega/\Xi$  ratios. 
Those discrepancies can be
traced back to the elementary UrQMD input.
A comparison with
figure~\ref{becattini}a) shows e.g. the underestimation of the 
elementary $\phi$-yield in proton-proton reactions by a factor of 2.

A thermal and chemical equilibrium model can be even used to 
fit the hadron ratios of the UrQMD calculation displayed
in the upper frame of figure~\ref{sau_ratios}.
The parameters
of the thermal model fit to the microscopic calculation in the  $y_{lab} = 3 \pm 0.5$
region (a detailed discussion of the rapidity dependence of the ratios is given below)
yields a temperature of $T=145$~MeV and a 
baryo-chemical potential of $\mu_B=165$~MeV.  
However, the assumption of global thermal and chemical equilibrium is not
justified:
Both, the discovery of directed collective flow of baryons and antiflow of mesons 
in Pb+Pb reactions at 160 GeV/nucleon
energies\cite{peitzmann97a} 
as well as transport model analyses, which show distinctly different
freeze-out times and radii for different hadron species \cite{sorge97a,soff97b},
indicate that
the yields and ratios result from a complex non-equilibrium
time evolution of the hadronic system. 
A thermal model fit to 
a nonequilibrium transport model (and to the data!) may therefore not seem meaningful.

\begin{figure}
\centerline{\psfig{figure=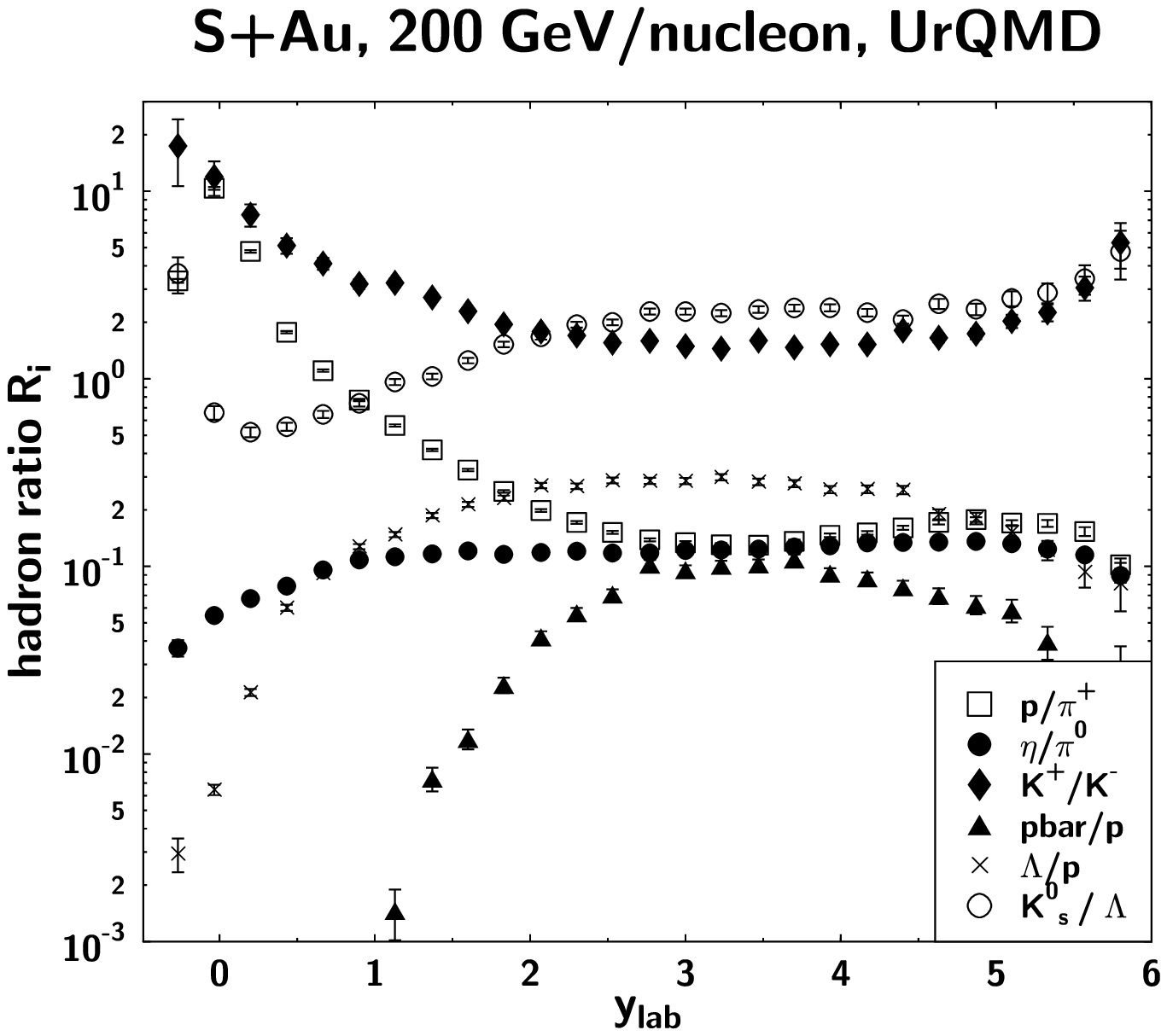,height=10cm}}
\caption{\label{y_ratios} Rapidity dependence of hadron ratios in
the UrQMD model for the system S+Au(W,Pb) at
CERN/SPS energies. The ratios vary by orders of magnitude, yielding
different $T$ and $\mu_B$ values  for different rapidity intervals.}
\end{figure}

Let's turn to the next question: 
To what extent do the hadron ratios depend on rapidity and
transverse momentum? How strong is their sensitivity to experimental
acceptance cuts?
The rapidity dependence of individual hadron ratios $R_i$
is shown in Figure~\ref{y_ratios}: 
The $p/\pi^+$, $\eta/\pi^0$, $K^+/K^-$, $\bar p/p$, $\Lambda/p$ and
$K^0_S/\Lambda$ ratios are plotted as a function of $y_{lab}$.
A strong dependence of the ratios $R_i$ on the rapidity is visible -- some
ratios, especially those involving (anti-)baryons, change
by orders of magnitude when going from target rapidity to mid-rapidity. 
The y-dependence 
is enhanced by the 
heavy target which leads to strong absorption of mesons and antibaryons.
The observed shapes of $R_i(y)$ are distinctly different from 
a fireball ansatz, incorporating additional longitudinal flow:
There, the ratios would also be symmetric with respect to the rapidity of the
central source.
A broad plateau would only be visible for ratios of particles with 
similar masses. When fitting a thermal model to data, one must
take this rapidity dependence into account and correct for different
experimental acceptances.

The large difference in the $K_0/\bar \Lambda$ ratio (as calculated by
UrQMD) visible between
figure~\ref{sau_ratios}a) and figure~\ref{sau_ratios}b) 
exemplifies the strong dependence of the hadron ratios 
on the experimental acceptances: While the experimental acceptance in rapidity
is similar to the cut employed in figure~\ref{sau_ratios}a), the 
additional cut in $p_t$, which has been performed in figure~\ref{sau_ratios}b),
changes the ratio by one order of magnitude.

\begin{figure}
\centerline{\psfig{figure=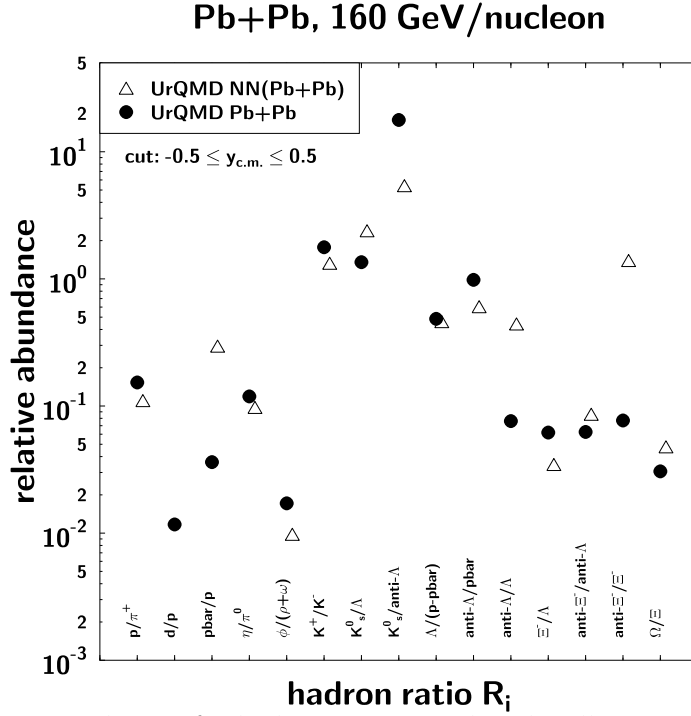,height=10cm}}
\caption{\label{pb_ratios} UrQMD prediction for hadron ratios in Pb+Pb
collisions at midrapidity (full circles). The ratios are compared to a
superposition of pp, pn and nn reactions with the isospin weight
of the Pb+Pb system (open triangles), i.e. a first collision approach. 
Especially in the sector of anti-baryons the ratios change by at least
one order of magnitude due to the large anti-baryon annihilation cross
section.}

\end{figure}

Figure~\ref{pb_ratios}, finally, shows the UrQMD 
prediction for the heavy system Pb+Pb. 
The ratios around midrapidity (full circles) 
are again compared to those stemming from 
isospin-weighted nucleon-nucleon calculation (open triangles). 
For this heavy system, rescattering effects are even larger than
in the S+Au case: Due to the large number of baryons around midrapidity,
antibaryon annihilation at midrapidity occurs more often and
therefore ratios involving antibaryons may be suppressed stronger
than in the S+Au case. Most prominent examples are (again) the
$\bar \Xi / \Xi$ (factor 20 suppression),
 $\bar p/p$ (factor 8
suppression) and the $K^0_S/\bar \Lambda$ (factor 3 enhancement) ratios.

In terms of absolute yields, the enhancement of particle 
production due to secondary interactions 
can be seen in figure~\ref{becattini}b). Here, a comparsion
between hadron yields in elementary p+p interactions at 
$\sqrt{s}=27$ GeV  and a respective 
Pb+Pb calculation with the yields scaled down by the relative 
number of participating
nucleon pairs, $A_{Pb}$, is plotted. 
The yields {\em per participating nucleon-pair} are enhanced by
factors of 2-10, especially those of resonances, 
such as the $\Delta_{1232}$, the $\rho$ 
or the $\Sigma^*$. Anti-resonance or -hyperon production 
is enhanced as well, their
large absorption cross sections, however, ``counter'' 
this enhancement (see the antiproton
suppression by a factor of 3). Thus, their yields
are similar to those found in proton-proton reactions.

Open questions, to be addressed in a forthcoming, more detailed publication,
include the dependence of $R_i$ on the azimuthal angle and the 
impact parameter.
Both should play a major role for baryon to antibaryon ratios, 
since those ratios are extremely sensitive to the phase-space 
distribution of baryonic matter \cite{jahns}.

Details in the treatment of the baryon-antibaryon annihilation cross section 
may have a large influence on the final yield of antiprotons and antihyperons:
If the proton-antiproton annihilation cross section 
as a function of $\sqrt{s}$ is used for all 
baryon-antibaryon annihilations, instead of rescaling the cross section 
to equivalent relative momenta,
the $\bar \Xi$ yield in central Pb+Pb reactions at 200 GeV/nucleon 
would be enhanced by a factor of 3. The $\bar p$ and $\bar Y$
yields would then be enhanced by 50\% and 25\%, respectively. 

A systematic study of different baryon to antibaryon ratios as 
functions of system size, impact parameter, transverse momentum and 
azimuthal angle may help to gain further insight into the 
antihyperon-nucleon and antihyperon-hyperon annihilation cross section.

In summary, 
ratios of hadronic abundances for $\sqrt{s} \sim 20$~GeV
have been analyzed within a microscopic transport model. A comparison
to data shows good agreement. Discrepancies can be found in the
$\phi/(\rho + \omega)$,
$K^0_S/\Lambda$ and $\Omega/\Xi$  ratios.
The resulting ratios have been compared to the primordial abundances from a 
cocktail of elementary $pp$, $pn$ and $nn$ interactions 
and then analyzed with respect to their dependence 
on secondary interactions and on rapidity. 
Their  strong dependence on rapidity casts doubt on the 
assumption of thermal and chemical equilibrium, which has been prevalent in
previous analyses. Finally, hadron ratios for the symmetric heavy system Pb+Pb
far from the elementary primordial nucleon-nucleon values are predicted.


\begin{references}
\bibitem{stoecker}
H.~St\"ocker, W.~Greiner, and W.~Scheid,
\newblock Z. Phys. {\bf A286}, 121 (1978). \\
D.~Hahn and H.~St\"ocker,
\newblock Nucl. Phys. {\bf A476}, 718 (1988).\\
D.~Hahn and H.~St\"ocker,
\newblock Nucl. Phys. {\bf A452}, 723 (1986).\\
H.~St\"ocker and W.~Greiner,
\newblock Phys. Rep. {\bf 137}, 277 (1986).
\bibitem{stock86a}
R.~Stock,
\newblock Phys. Rep. {\bf 135}, 261 (1986).
\bibitem{rafelski}
J.~Rafelski and B.~M\"uller,
\newblock Phys. Rev. Lett. {\bf 48}, 1066 (1982).\\
J.~Rafelski,
\newblock Phys. Rep. {\bf 88}, 331 (1982).\\
P.~Koch, B.~M\"uller, and J.~Rafelski,
\newblock Phys. Rep. {\bf 142}, 167 (1986).
\bibitem{rafelski2}
J.~Letessier, A.~Tounsi, U.~Heinz, J.~Sollfrank, and J.~Rafelski,
\newblock Phys. Rev. Lett. {\bf 70}, 3530 (1993).\\
J.~Letessier, J.~Rafelski, and A.~Tounsi,
\newblock Phys. Lett. {\bf B321}, 394 (1994).\\
J.~Rafelski and M.~Danos,
\newblock Phys. Rev. {\bf C50}, 1684 (1994).\\
J.~Sollfrank, M.~Gadzicki, U.~Heinz, and J.~Rafelski,
\newblock Z. Phys. {\bf C61}, 659 (1994).
\bibitem{cleymans}
J.~Cleymans, M. I. Gorenstein, J. Stalnacke and E. Suhonen,
\newblock Phys. Scripta {\bf 48}, 277 (1993).\\
J.~Cleymans and H.~Satz,
\newblock Z. Phys. {\bf C57}, 135 (1993).\\
J. Cleymans, D. Elliott, H. Satz, R.L. Thews,
\newblock Z. Phys. {\bf C74},319 (1997) 
\bibitem{braun-munzinger}
P.~Braun-Munzinger, J.~Stachel, J.~P. Wessels, and N.~Xu,
\newblock Phys. Lett. {\bf B344}, 43 (1995), nucl-th/9410026 and
\newblock Phys. Lett. {\bf B365}, 1 (1996), nucl-th/9508020. \\
P.~Braun-Munzinger and J.~Stachel, 
\newblock  Nucl. Phys. {\bf A606}, 320 (1996). 
\bibitem{spieles97a}
C.~Spieles, H.~St\"ocker, and C.~Greiner,
\newblock Z. Phys. {\bf C}, in print (1997), nucl-th/9704008.\\
C.~Greiner and H.~St\"ocker.,
\newblock Phys. Rev. {\bf D44}, 3517 (1992).
\bibitem{exis_ags}
S.~E. Eiseman {\em et~al.},
\newblock Phys. Lett. {\bf B297}, 44 (1992).\\
J.~Barrette {\em et~al.},
\newblock Z. Phys. {\bf C59}, 211 (1993).\\
T.~Abbott {\em et~al.},
\newblock Phys. Rev. {\bf C50}, 1024 (1994).\\
G.~S.~F. Stephans {\em et~al.},
\newblock Nucl. Phys. {\bf A566}, 269c (1994).\\
S.~E. Eiseman {\em et~al.},
\newblock Phys. Lett. {\bf B325}, 322 (1994).\\
J.~Barrette {\em et~al.},
\newblock Phys. Lett. {\bf B351}, 93 (1995).
\bibitem{exis_sps}
E.~Andersen {\em et~al.},
\newblock Phys. Lett. {\bf B294}, 127 (1992).\\
E.~Andersen {\em et~al.},
\newblock Phys. Lett. {\bf B327}, 433 (1994).\\
M.~Murray {\em et~al.},
\newblock Nucl. Phys. {\bf A566}, 589c (1994).\\
J.~T. Mitchell,
\newblock Nucl. Phys. {\bf A566}, 415c (1994).\\
M.~Gadzicki {\em et~al.},
\newblock Nucl. Phys. {\bf A590}, 197c (1995).\\
S.~Abatzis {\em et~al.},
\newblock Phys. Lett. {\bf B316}, 615 (1993).\\
D.~D. Bari {\em et~al.},
\newblock Nucl. Phys. {\bf A590}, 307c (1995).
\bibitem{becattini97a}
F.~Becattini and U.~Heinz,
\newblock Z. Phys. {\bf C76}, 269 (1997), hep-ph/9702274.
\bibitem{barz88a}
H.~W. Barz, B.~L. Friman, J.~Knoll, and H.~Schultz,
\newblock Nucl. Phys. {\bf A484}, 661 (1988).
\bibitem{uqmdref}
S.~A.~Bass, M.~Belkacem, M.~Bleicher, M.~Brandstetter, C.~Ernst, 
L.~Gerland, C.~Hartnack, S.~Hofmann, J.~Konopka, G.~Mao, L.~Neise, S.~Soff,
C.~Spieles, H.~Weber, N.~Amelin, J.~Aichelin, H.~St\"ocker  and W.~Greiner,
\newblock to appear in Progr. Part. Nucl. Physics Vol. {\bf 41} (1998).\\
M.~Bleicher, C.~Spieles, L.~Gerland, S.~A.~Bass, M.~Belkacem, M.~Brandstetter, C.~Ernst, 
S.~Hofmann, J.~Konopka, G.~Mao, L.~Neise, S.~Soff,
H.~Weber H. St\"ocker  and W. Greiner,
\newblock to be sumitted to Phys.~Rev.~{\bf C}
\bibitem{peilert88a}
G.~Peilert, A.~Rosenhauer, H.~St\"ocker, W.~Greiner, and J.~Aichelin,
\newblock Modern Physics Letters {\bf A3}, 459 (1988).
\bibitem{hartnack89b}
C.~Hartnack {\em et~al.},
\newblock Nucl. Phys. {\bf A495}, 303 (1989).
\bibitem{aichelin91a}
J.~Aichelin,
\newblock Phys. Rep. {\bf 202}, 233 (1991).
\bibitem{sorge89a}
H.~Sorge, H.~St\"ocker, and W.~Greiner,
\newblock Annals of Physics {\bf 192}, 266 (1989).
\bibitem{lehmann95a}
E.~Lehmann, R.~Puri, A.~Faessler, G.~Batko, and S.~Huang,
\newblock Phys. Rev. {\bf C51}, 2113 (1995).
\bibitem{pop95a}
V. Topor Pop, M. Gyulassy, X.N. Wang, A. Andrighetto, M. Morando, 
F. Pellegrini, R.A. Ricci and G. Segato.
\newblock Phys. Rev. {\bf C52}, 1618 (1995). 
\bibitem{peitzmann97a}
T.~Peitzmann {\em et~al.},
\newblock Proceedings of the International Workshop on Gross Properties of
  Nuclei and Nuclear Excitation XXV, {\em QCD Phase Transitions}, Hirschegg,
  Kleinwalsertal (Austria), January 1997 .\\
H.~Gutbrod,
\newblock private communication.
\bibitem{sorge97a}
H.~Sorge,
\newblock Phys. Rev. Lett. {\bf 78}, 2309 (1997).
\bibitem{soff97b}
M. Bleicher, S.A. Bass,  M. Belkacem, J. Brachmann, M. Brandstetter,
C. Ernst, L. Gerland, J. Konopka, S. Soff, C. Spieles, H. Weber, H. St\"ocker
and W. Greiner.
\newblock Proceedings of the 35th International Winter Meeting on Nuclear Physics, 
Bormio, Italy, 26 Jan - 1 Feb 1997. 
e-Print Archive: nucl-th/9704065 \\ 
S.~Soff, S.A. Bass, S. Schramm, M. Bleicher, 
J. Konopka, C. Spieles, H. Weber, H. St\"ocker
and W. Greiner.
\newblock manuscript in preparation.
\bibitem{jahns}
A. Jahns, C. Spieles, H. Sorge, H. Stocker, W. Greiner,
\newblock Phys. Rev. Lett.{\bf 72}, 3464 (1994). 
\end{references}

\end{document}